\documentclass[usenatbib]{mn2e}
\pdfoutput=1
\usepackage{mathtools}
\usepackage{graphicx}
\usepackage{enumerate}
\usepackage{amsmath}
\usepackage{breqn}
\usepackage{epstopdf}

\usepackage{array}
\usepackage{amssymb}
\usepackage{multirow}
\usepackage{caption}
\usepackage{longtable}
\usepackage{supertabular}
\usepackage{lineno}

\graphicspath{{./pics/}}
\begin{document}
\title[Chameleon gravity with clusters]{The \textit{XMM} Cluster Survey: Testing chameleon gravity using the profiles of clusters}
\author[H. Wilcox et al.]
{Harry Wilcox$^1$\thanks{harry.wilcox@port.ac.uk}, David Bacon$^1$, Robert C. Nichol$^1$, Philip J. Rooney$^2$, \newauthor Ayumu Terukina$^3$, A. Kathy Romer$^2$, Kazuya Koyama$^1$, Gong-Bo Zhao$^{4,1}$, \newauthor Ross Hood$^5$, Robert G. Mann$^5$, Matt Hilton$^6$, Maria Manolopoulou$^5$, \newauthor Martin Sahl\'en$^{7}$, Chris A. Collins$^8$, Andrew R. Liddle$^5$, Julian A. Mayers$^2$, \newauthor Nicola Mehrtens$^{9,10}$, Christopher J. Miller$^{11}$, John P. Stott$^{12}$, Pedro T. P. Viana$^{13,14}$\\
$^1$ Institute of Cosmology and Gravitation, University of Portsmouth, Dennis Sciama Building, Portsmouth, PO1 3FX, UK \\
$^2$ Astronomy Centre, University of Sussex, Falmer, Brighton, BN1 9QH, UK \\
$^3$ Department of Physical Science, Hiroshima University, Higashi-Hiroshima 739-8526, Japan \\
$^4$ National Astronomy Observatories, Chinese Academy
of Science, Beijing, 100012, P.R. China \\
$^5$ Institute for Astronomy, University of Edinburgh, Royal Observatory, Blackford Hill, Edinburgh, EH9 3HJ, UK\\
$^6$ Astrophysics \& Cosmology Research Unit, School of Mathematics, Statistics \& Computer Science, University of KwaZulu-Natal, \\ Westville Campus, Durban 4041, South Africa \\
$^{7}$ BIPAC, Department of Physics, University of Oxford, Denys Wilkinson Building, 1 Keble Road, Oxford OX1 3RH, UK \\
$^8$ Astrophysics Research Institute, Liverpool John Moores University, IC2, Liverpool Science Park, Brownlow Hill, Liverpool, L5 3AF \\
$^9$ George P. and Cynthia Woods Mitchell Institute for Fundamental Physics and Astronomy, Texas A\&M University, College Station, \\ TX, 77843-4242 USA \\
$^{10}$ Department of Physics and Astronomy, Texas A\&M University, College Station, TX, 77843-4242 USA \\
$^{11}$ Astronomy Department, University of Michigan, Ann Arbor, MI 48109, USA \\
$^{12}$ Sub-department of Astrophysics, Department of Physics, University of Oxford, Denys Wilkinson Building, Keble Road, \\ Oxford OX1 3RH, UK \\
$^{13}$ Instituto de Astrof\'{\i}sica e Ci\^{e}ncias do Espa\c{c}o, Universidade do Porto, CAUP, Rua das Estrelas, 4150-762 Porto, Portugal \\
$^{14}$ Departamento de F\'isica e Astronomia, Faculdade de Ci\^encias, Universidade do Porto, Rua do Campo Alegre, 687, \\ 4169-007 Porto, Portugal}
\maketitle
\begin{abstract}
The chameleon gravity model postulates the existence of a scalar field that couples with matter to mediate a fifth force.  If it exists, this fifth force would influence the hot X-ray emitting gas filling the potential wells of galaxy clusters. However, it would not influence the clusters’ weak lensing signal. Therefore, by comparing X-ray and weak lensing profiles, one can place upper limits on the strength of a fifth force. This technique has been attempted before using a single, nearby cluster (Coma, $z=0.02$, \citealt{2014JCAP...04..013T}). Here we apply the technique to the stacked profiles of 58 clusters at higher redshifts ($0.1<z<1.2$ ), including 12 new to the literature, using  X-ray data from the XMM Cluster Survey (XCS) and weak lensing data from the Canada France Hawaii Telescope Lensing Survey (CFHTLenS). Using a multi-parameter MCMC analysis, we constrain the two chameleon gravity parameters ($\beta$ and $\phi_{\infty}$). Our fits are consistent with general relativity, not requiring a fifth force. In the special case of $f(R)$ gravity (where $\beta = \sqrt{1/6}$), we set an upper limit on the background field amplitude today of $|f_{\rm{R0}}| < 6 \times 10^{-5}$ (95\% CL). This is one of the strongest constraints to date on $|f_{\rm{R0}}|$ on cosmological scales. We hope to improve this constraint in future by extending the study to hundreds of clusters using data from the Dark Energy Survey.

\end{abstract}

\section{Introduction}
An accepted explanation for the accelerated expansion of the late-time Universe (\citealt{1998AJ....116.1009R}, \citealt{1999ApJ...517..565P}) is to modify the Einstein equation, either by adding a component to the energy-momentum tensor via dark energy, or to the Einstein tensor via a modification to gravity (\citealt{1983ApJ...270..365M}, \citealt{2012PhR...513....1C}).  The latter often involves the introduction of a scalar field coupled to the matter components of the Universe, giving rise to a fifth force of the same order of magnitude as gravity \citep{2013ApJ...779...39J}.  Through a variety of experiments and astronomical observations, this fifth force has been demonstrated to be negligible at terrestrial and solar system densities \citep{2012CQGra..29r4002W}.  Therefore, if a fifth force does exist it must be it must be suppressed, or ``screened'', in high density regions and only take effect in low density regions.

One model with such a screening is the chameleon mechanism \citep{Khoury:2003aq}. In this approach, the scalar field coupling strength is sensitive to the depth of the local gravitational potential.  In regions with a large potential well this screening suppresses the fifth force and gravity behaves as predicted by general relativity.  However when the potential becomes small, the fifth force is unsuppressed and gravity  becomes ``modified'' compared to general relativity \citep{2014AnP...526..259L}.

By definition, the chameleon field satisfies
\begin{equation}
\nabla^{2} \phi = V_{,\phi} + \frac{\beta}{M_{\rm{Pl}}} \rho,
\label{eq:cham_field}
\end{equation}
\citep{Khoury:2003aq}, where $V$ is the potential of the scalar field, $\beta$ is the coupling between matter and the scalar field, $\phi$ gives the position dependent  screening efficiency, $M_{\text{Pl}}$ is the Planck mass and $\rho$ is the matter density.  This leads to the chameleon fifth force of
\begin{equation}
F_{\phi} = - \frac{\beta}{M_{\rm{Pl}}} \nabla \phi.
\label{eq:5thforce}
\end{equation}

There is a particular set of gravity models, known as $f(R)$ models \citep{1970MNRAS.150....1B} which exhibit a chameleon, where the strength of the fifth force (parameterised by $\beta$ in Equation \ref{eq:cham_field}) has a fixed value $\beta=\sqrt{1/6}$. This force arises from adding a scalar function $f(R)$ to the Ricci scalar in the Einstein-Hilbert action (\citealt{Capozziello:2002rd}, \citealt{Nojiri:2003ft}).  These models can reproduce observed late time acceleration of the Universe whilst still suppressing the fifth force in high-density environments, such as the Solar System \citep{2007PhRvD..75l4014C}.  These $f(R)$ models possess an extra scalar degree of freedom, $f_{\rm{R}} = \text{d}f / \text{d}R$, where the value at the current epoch is  $|f_{\rm{R0}}|$ \citep{2010RvMP...82..451S}.  Then $f(R)$ gravity can be related to $\phi_{\infty}$, ($\phi$ in Equation \ref{eq:5thforce} at infinity) via the relation \citep{2014arXiv1407.0059J}
\begin{equation}
f_{\rm{R}}(z) = - \sqrt{\frac{2}{3}}\frac{\phi_{\infty}}{M_{\text{Pl}}}.
\label{eq:f_R}
\end{equation}

\cite{Hu:2007nk} provide theoretical arguments showing that for general relativity to be preserved at parsec scales within the Solar System, then $|f_{\rm{R0}}| < 10^{-6}$.  At kiloparsec scales \citealt{2013ApJ...779...39J} constrained $|f_{\rm{R0}}| < 5 \times 10^{-7}$ in dwarf galaxies.  On megaparsec and larger scales, \cite{2014PhRvD..90d3513R} used the Cosmic Microwave Background to measure $|f_{\rm{R0}}| < 10^{-3}$.  Also on large scales \cite{2011PThPS.190..179R}, \cite{Ferraro:2010gh}, \cite{2014arXiv1412.0133C} used the abundance of galaxy clusters to constrain $|f_{\rm{R0}}|$, e.g. \cite{2014arXiv1412.0133C} measured (under the assumption of $n=1$), $|f_{\rm{R0}}| < 2.6 \times 10^{-5}$.

In this paper, we also use clusters of galaxies to constrain $|f_{\rm{R0}}|$ on megaparsec scales.  However, unlike \cite{2011PThPS.190..179R}, \cite{Ferraro:2010gh}, \cite{2014arXiv1412.0133C}, we use cluster profiles, rather than abundances to do so.  The hypothesis is that a fifth force would be screened in the dense cluster cores, but not in the rarefied cluster outskirts (\citealt{Burikham:2011zr}, \citealt{2012PhRvD..85l4054L}). The majority of baryonic matter in a cluster is ionised gas that has been pressure-heated to temperatures in excess of $10^7$K (\citealt{1971ApJ...167L..81G}, \citealt{2004oee..symp..422L}), leading to the emission of X-rays via thermal bremsstrahlung radiation (\citealt{1978ApJ...224....1J}, \citealt{2009xecg.book.....S}). The gas can also be observed indirectly through its influence on the cosmic background radiation, via the so-called Sunyaev-Zel'dovich (SZ) effect \citep{1980ARA&A..18..537S}. 

By measuring the properties of this X-ray gas we are able to infer, under the assumption of hydrostatic equilibrium, the cluster mass and density from its X-ray surface brightness or SZ effect profiles (\citealt{2002ApJ...567..716R}, \citealt{2014arXiv1410.8769K}).  In a chameleon gravity model, the intracluster gas would feel the fifth force in addition to gravity in the cluster outskirts, i.e. the gas will be slightly more compact and the temperature boosted \citep{2014MNRAS.440..833A}, compared to the influence of general relativity alone.  

By contrast, weak gravitational lensing is dependent only upon the gravitational deflection of light by matter along the line of sight, therefore providing a technique to measure the underlying mass distribution without assuming hydrostatic equilibrium. Crucially for this study, the fifth force would not modify the deflection of light through the cluster (compared to general relativity) because the scalar chameleon field is coupled to the trace of the energy-momentum tensor \citep{2009PhRvD..80j4002H}. Therefore, we can search for evidence of a fifth force by comparing the X-ray surface brightness, and/or SZ effect, profiles of clusters with their gravitational lensing shear profiles (\citealt{1986Natur.322..804O}, \citealt{2012PhRvD..86j3503T}). 

\cite{2014JCAP...04..013T} used this approach to constrain $f(R)$ gravity models using a combination of lensing shear, X-ray surface brightness, X-ray temperature and Sunyaev-Zel'dovich profiles for the Coma cluster (a massive cluster at $z=0.02$).  Combining these measurements, they performed an MCMC analysis of the parameter space describing the cluster profiles in the modified gravity regime.  Under the assumption of hydrostatic equilibrium, they obtained constraints of $|f_{\rm{R0}}| < 6 \times 10^{-5}$.  They also examined the assumption of hydrostatic equilibrium, and concluded that any contribution of non-thermal pressure was small compared to the reconstructed mass. 

The Coma cluster is at low redshift meaning its weak lensing shear signal is low.  Moreover it is known to have non-spherical geometry (\citealt{1987ApJ...317..653F}, \citealt{1992A&A...259L..31B}, \citealt{1996ApJ...458..435C}).  These factors motivate us to apply the \cite{2014JCAP...04..013T} method to many more clusters at higher redshifts, allowing for a higher signal to noise weak lensing shear profile and an averaging out of non-spherical cluster shapes.  We do this by comparing stacked X-ray surface brightness and shear profiles of 58 X-ray selected clusters. We utilise high quality weak lensing data from the  Canada France Hawaii Telescope Lensing Survey (CFHTLenS, \citealt{2012MNRAS.427..146H}, \citealt{2013MNRAS.433.2545E}), and X-ray observations from the XMM Cluster Survey (XCS, \citealt{1999astro.ph.11499R}, \citealt{2011MNRAS.418...14L}, \citealt{2012MNRAS.423.1024M}).  We also investigate the \cite{2014JCAP...04..013T} conclusion that deviations from hydrostatic equilibrium do not invalidate the chameleon gravity test.

In Section \ref{sec:Theory} we review the underlying theoretical background. In Section \ref{sec:Methods} we describe the development of the cluster sample used in the analysis, and the Markov chain Monte Carlo (MCMC) methods used to simultaneously fit the X-ray surface brightness and weak lensing profiles. In Section \ref{sec:results} we discuss our results and the implications of our results in the framework of $f(R)$ gravity models. In Section \ref{sec:Discussion} we discuss the influence of cluster environment and of our assumption of hydrostatic equilibrium.  In Section \ref{sec:Conclusions} we present our conclusions.   Throughout this paper we use a 95\% confidence level when quoting upper limits, adopt a cosmology with $\Omega_{\rm{m}}=0.27$, $\Omega_{\Lambda}=0.73$ and $H_{\rm{0}} = 70$ km s$^{-1}$ Mpc$^{-1}$.

\section{Theoretical background}
\label{sec:Theory}

In this study, we adopt the Navarro-Frenk-White (NFW; \citealt{1996ApJ...462..563N}) model for the dark matter halo mass distribution:
\begin{equation}
\rho(r)=\frac{\rho_{\rm{c}}\delta_{\rm{c}}}{\frac{r}{r_{\rm{s}}}(1+\frac{r}{r_{\rm{s}}})^{2}},
\label{eq:NFW1}
\end{equation}
where $r$ here and throughout is the radial distance from the halo centre, $\rho_{\rm{c}}=3H^{2}(z)/8\pi G$ is the critical density at a given redshift, $H(z)$ is the Hubble parameter at a given redshift, $G$ is Newton's Gravitational Constant, $\delta_{\rm{c}}$ is the characteristic overdensity, given by
\begin{equation}
\delta_{\rm{c}} = \frac{200}{3} \frac{c^{3}}{\ln(1+c)-c/(1+c)}, 
\label{eq:NFW2}
\end{equation}
where $c$ is a dimensionless concentration parameter and $r_{\rm{s}}$ is the scale radius given by
\begin{equation}
r_{\rm{s}} = \frac{1}{c} \left( \frac{3 M_{\rm{200}}}{4 \pi \rho_{\rm{c}} \delta_{\rm{c}}} \right)^{1/3},
\label{eq:NFW3}
\end{equation}
where $M_{\rm{200}}$ is the mass enclosed by $r_{\rm{200}}$, the radius at which the dark matter halos average density is two hundred times the critical density, 
\begin{equation}
M(<r_{200})=4 \pi \delta_{\rm{c}} \rho_{\rm{c}} r_{\rm{s}}^{3} \bigg ( \ln(1+c) - \frac{c}{1+c} \bigg ).
\label{eq:NFW4}
\end{equation}

The NFW profile described in Equation \ref{eq:NFW4} is well supported by N-body simulations of $\Lambda \text{CDM}$, but it is not immediately obvious that this profile would pertain to cluster profiles in the $f(R)$ regime.  However it has been shown (\citealt{2012PhRvD..85l4054L}, \citealt{Moran21032015}) that the NFW profile is able to provide fits to both modified gravity and concordance cosmology that are equally good, sharing the same $\chi^{2}$.  It should be noted that the simulations in \cite{2012PhRvD..85l4054L} were generated using a fixed $\beta = \sqrt{1/6}$, as opposed to the general chameleon gravity model investigated here.  However as we are probing a $\beta$ range around this value, we expect any modifications to the profiles to be similar, suggesting the suitability of the NFW profile.  Further checks using hydrodynamical simulations of modified gravity models would allow this assumption to be verified.

We adopt the \cite{2014JCAP...04..013T} approach describing the chameleon mechanism using three parameters.  The first of these, $\beta$, is the coupling between matter and the scalar field (see Equation \ref{eq:cham_field}). The second, $\phi_{\infty}$, describes the position dependent screening efficiency.  The third, $r_{\rm{crit}}$, is a critical radius, i.e. the distance from the dark matter halo centre at which the screening mechanism takes effect \citep{2012PhRvD..86j3503T},
\begin{equation}
r_{\rm{crit}} = \frac{\beta \rho_{\rm{s}} r^{3}_{\rm{s}}}{M_{\rm{Pl}} \phi_{\rm{\infty}}} - r_{\rm{s}},
\label{eq:rc}
\end{equation}
where $\rho_{\rm{s}}$ is the density at this radius.


\cite{2012PhRvD..86j3503T} showed the hydrostatic equilibrium equation in the presence of a fifth force (Equation \ref{eq:5thforce}) is:
\begin{equation}
\frac{1}{\rho_{\rm{gas}}(r)} \frac{dP_{\rm{gas}}(r)}{dr} = - \frac{GM(<r)}{r^{2}} - \frac{\beta}{M_{\rm{Pl}}} \nabla \phi,
\label{eq:hydro_equil}
\end{equation}
where $\rho_{\rm{gas}}$ is the gas density,  $M$ the total mass within a radius $r$, and $P_{\rm{gas}}$ is the electron pressure.  

In an ideal cluster, i.e. one that is isolated, isothermal, and spherical, this total pressure is felt by the electrons and ions in the ionised intracluster plasma, so that $P_{\rm{gas}}=n_{\rm{e}}kT$, where $n_{\rm{e}}$ is the electron number density, and $T$ is the electron temperature.  By adopting the standard {\it beta-model}\footnote{The {\it beta} in this model is not the same as the $\beta$ in Equation \ref{eq:cham_field}.} electron density profile (e.g. \citealt{1978A&A....70..677C}), we can integrate Equation \ref{eq:hydro_equil} to give
\begin{align}
P_{\rm{e}}(r) = P_{\rm{e,0}} + \mu m_{\rm{p}} \int^{r}_{\rm{0}} n_{e}(r) \nonumber \\ \left( -\frac{GM(<r)}{r^{2}} - \frac{\beta}{M_{\rm{Pl}}} \frac{d \phi(r)}{dr} \right) dr,
\label{eq:Pintegral}
\end{align}
where $P_{\rm{e,0}}$ is the electron gas pressure at $r=0$, given by $P_{\rm{e,0}}=n_{\rm{e,0}}kT$ and $n_{\rm{e,0}}=5n_{\rm{0}}/(2+\mu)$ and $M(<r)$, the halo mass. The integral of Equation \ref{eq:Pintegral} can be re-expressed  in terms of a projected X-ray surface brightness $S_{\rm{B}}(r)$ using the temperature and 
electron density dependent cooling function (see Section \ref{sec:Measure_X_ray}), 
\begin{equation}
S_{\rm{B}}(r_{\perp}) = \frac{1}{4 \pi (1+z)^{4}} \int n^{2}_{\rm{e}} \left( \sqrt{r_{\perp}^{2} + z^{2}} \right) \lambda_{c}(T_{\rm{gas}})dz,
\label{eq:SBprofile}
\end{equation}
where $r_{\perp}$ is the projected distance from the cluster centre and $z$ the cluster redshift. This is the expression we fit to when comparing stacked X-ray cluster profiles to the chameleon model (Section \ref{sec:mcmc}).

The expression used to fit the weak lensing shear profiles (under the assumption of an underlying NFW profile) for comparison is given in \citealt{0004-637X-534-1-34}.

To recap, our method makes the following assumptions: that modifications to general relativity include a chameleon screening mechanism and can be described by Equation \ref{eq:cham_field}; that dark matter halos follow an NFW profile (Equation \ref{eq:NFW1}); that a fifth force can be included in the hydrostatic equilibrium expression according to Equation \ref{eq:hydro_equil}; that clusters of galaxies are isolated, isothermal, and spherical (which in turn implies that the clusters are in hydrostatic equilibrium, have an electron number density that follows a {\it beta-model}, and their X-ray emission can be predicted from a thermal cooling function); and that the weak lensing shear profiles of clusters are given in \citealt{0004-637X-534-1-34}. We discuss the impact of some of these assumptions in Section \ref{sec:Discussion}.


\section{Methods}
\label{sec:Methods}
\subsection{Compiling the X-ray Cluster Sample}

In this paper we used public weak lensing data (galaxy ellipticities and photometric redshifts) provided by the Canada France Hawaii Lensing Survey (CFHTLenS, \citealt{2012MNRAS.427..146H}). The CFHTLenS covers 154 sq. deg with high quality shape measurements.  The galaxy ellipticities were generated by the CFHTLenS team using the {\tt THELI} \citep{2013MNRAS.433.2545E} and {\tt lensfit} \citep{2013MNRAS.429.2858M} routines.  Photometric redshifts were produced using PSF-matched photometry to an accuracy of $0.04(1+z)$ with a 4\% catastrophic outlier rate \citep{2012MNRAS.421.2355H}.

We also used public X-ray data taken from the {\it XMM-Newton} archive and have collated a sample of X-ray clusters in the CFHTLenS region using pipelines developed for the XMM Cluster Survey (XCS, \citealt{2011MNRAS.418...14L}). First we determined which of the {\it XMM} observations overlapped with the CFHTLenS fields. We then used the XCS pipelines to carry out the following tasks in an automated manner: cleaning the event lists of background flares; creating detector and exposure images; producing duplicate free source lists; and identifying extended X-ray sources. A total of 348 extended {\it XMM} sources, with more than 100 background subtracted photon counts, were located in the CFHTLenS fields, although 44 were close to the edge of the {\it XMM} field of view and were not considered further (please see \citealt{2011MNRAS.418...14L} for the relevant, XCS specific, definition of source counts).

The majority of these sources were not included in the XCS first data release (XCS-DR1, \citealt{2012MNRAS.423.1024M}).  This meant that candidate identification needed to be carried out before the sources could be used in our study.   This process is non trivial: as shown in \citealt{2012MNRAS.423.1024M},  a large fraction of XCS extended sources (especially those with fewer than 300 counts) are either hard to confirm as clusters -- because the available imaging is not deep enough -- or are associated with other types of X-ray source. Therefore, for this paper, we have taken a conservative approach and only included {\it XMM} extended sources in our study if they correspond to an over density of galaxies in false colour images produced using the CFHTLenS cutout service\footnote{http://www.cadc-ccda.hia-iha.nrc-cnrc.gc.ca/community/CFHTLens/cutout.html}. One hundred and eighty six sources were excluded from the study as a result. These were excluded for several different reasons: there was no optical data as the cluster sat in a masked region of the CFHTLenS footprint; there was a bright star or galaxy lying close to the cluster centre that was obscuring it; or the optical image resembled an AGN rather than a cluster. The coordinates of the remaining one hundred and nineteen can be found in Table \ref{table:sources}.

As our analysis required information about the distance to the cluster, a further 37 sources were excluded from the study because redshifts were not available at the time of writing. These are flagged with a 2 in Table \ref{table:sources}. The majority (63 of 82) of the redshifts we used came from the new Gaussian mixture model redshift estimator described in detail in \citealt{Hood.paper}.  We also used 18 redshifts taken from {\tt NED\footnote{The NASA/IPAC Extragalactic Database (NED) is operated by the Jet Propulsion Laboratory, California Institute of Technology, under contract with the National Aeronautics and Space Administration.}} and 3 from \cite{2014MNRAS.439.3755F}.

We judged these remaining 82 {\it XMM} extended sources in the CFHTLenS region to be confirmed clusters and ran them through the XSPEC based XCS spectral pipeline. We determined X-ray temperatures when the signal to noise was sufficient. This produced X-ray temperatures of 58 of these clusters which form our final sample, including 12 clusters new to the literature, the other 23 clusters were excluded from the analysis and are flagged with a 3 in Table \ref{table:sources}. The details of this pipeline can be found in \citealt{2011MNRAS.418...14L}. These 58 clusters with measured temperatures span the redshift range $0.1<z<1.2$ (median $z=0.33$) and temperature range  $0.2<T_{\rm{x}}<8$~keV (median $T_{\rm{x}}=2.3$~keV).  A selection of these new to the literature clusters, along with several clusters that were optically confirmed but excluded due to a lack of redshift, are shown in Figure \ref{fig:clust_pics}.

\subsection{Making stacked X-ray Surface Brightness Profiles}
\label{sec:Measure_X_ray}


Our analysis involves stacking multiple different {\it XMM} observations of our 58 clusters, in order to build up signal to noise in the outer parts of the ensemble cluster profile. This process needs to account for the following complexities: Most of the 58 clusters were covered by more than one {\it XMM} observation.  Each of these observations has different background properties and flare corrected exposure times.  The X-ray telescope comprises of three cameras that operate simultaneously (mos1, mos2, pn), so most {\it XMM} observations comprise of three separate images with different, energy dependant sensitivities. The clusters all have different energy spectra, because, even if one ignores non thermal processes, they have different X-ray temperatures, redshifts, and line of sight absorbing column densities. Therefore, for each cluster, we have to calculate, using XSPEC, camera specific count rate to luminosity conversion factors for each {\it XMM} observation that it falls in. We then, for a given cluster, take the photon count images generated by the XCS pipeline, divide these by the respective exposure map, and multiply by the cluster dependent conversion factor.  This allows us to combine all the images for that cluster in a self consistent manner.

To produce a single stack, we first re-scaled the 58 combined images of individual clusters to a standard projected size. For this we estimated $M_{\rm{500}}$, the mass enclosed within a sphere at which the average density is 500 times the critical density, using the prescription described in \cite{2009MNRAS.397..577S}.  A conversion between $M_{\rm{500}}$ and $M_{\rm{200}}$ was made following the formulae derived in \cite{2003ApJ...584..702H}, where we assume $c=5$.  This is an accurate description of the typical density profiles in clusters \citep{2005bmri.conf...77A} and is consistent with the findings of \cite{2014arXiv1410.8769K} in the CFHTLenS region. Using the $M_{\rm{200}}$ values we calculated the radius at which the average density is two hundred times the critical density, $r_{\rm{200}}$, following the method in \cite{2008A&A...487..431C}.  The 58 stacked images could then be rescaled using linear interpolation to a common 500 by 500 pixel format, so that they each had an $r_{\rm{200}}$ radius of 125 pixels. Each of these 500 by 500 images were centred on the source centroid as determined by XCS.

We re-scaled the individual cluster images by the overall amplitude of their X-ray surface brightness, as adding clusters over a range of different masses and luminosities would result in significant off-diagonal elements in the covariance matrix of the final stacked profile. Therefore, we calculate the mean value of the X-ray surface brightness profile for each cluster, and re-scale individual cluster surface brightness maps by this value (we found that using the median value instead of the mean gave similar results). A final stacked surface brightness map of the 58 individual clusters is then produced by taking the mean value for each pixel across all these maps. This re-scaling of the amplitudes is permitted as our constraints on modified gravity parameters focus on the shape of the cluster profiles; we marginalise over the amplitudes of the stacked X-ray surface brightness profiles in Section 4.  The error covariance matrix of the stacked profile was then measured directly.

\subsection{Making Stacked Weak Lensing Profiles}

\label{sec:measure_lensing}
\begin{figure*}
\centerline{\includegraphics[width=21cm]{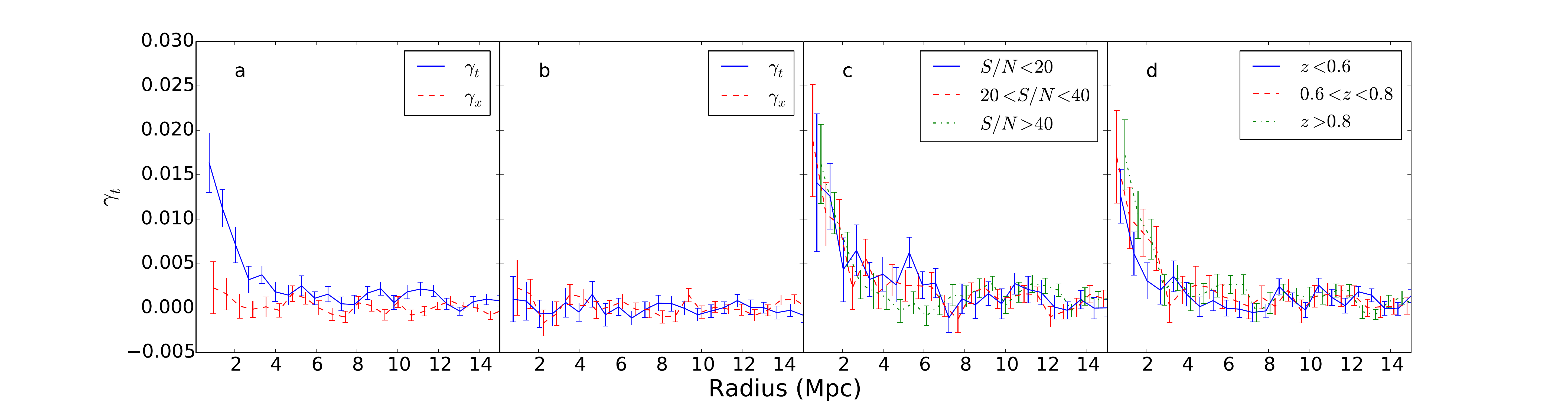}}
\caption{Tests around the 58 CFHTLenS stacked cluster; details are provided in the text.  1a: Tangential and cross shear. 1b: Tangential and cross shear around 58 stacked random points.  1c: Tangential shear for three different signal to noise bins.  1d: Tangential shear for three different redshift cuts.}
\label{fig:null_tests}
\end{figure*}

We outline here the procedure used to obtain the stacked cluster shear profile, $\gamma_{\rm{t}}$, using source galaxies from CFHTLenS.  The CFHTLenS catalogue provides measurements of both ellipticity components ($e_{\rm{1}}$ and  $e_{\rm{2}}$), as well as photometric redshifts for each source galaxy. Before shears can be derived from these quantities, small multiplicative and additive corrections ($m$ and $c_{\rm{2}}$) must be applied, derived from the dataset.  We calculate $c_{\rm{2}}$ and $m$ for each galaxy as a function of size and signal to noise (using Equation 17 and 19 in \citealt{2012MNRAS.427..146H}).  Each galaxy was weighted with the CFHTLenS catalogue WEIGHT parameter and calibrated by
\begin{equation}
e_{\rm{int,i}} = \frac{e_{\rm{i}} - c_{\rm{2,i}}}{1 + \bar{m}},
\end{equation}
where $c_{\rm{2}}$ was applied on a galaxy by galaxy basis and $\bar{m}$ is a summation of $1+m$ for each galaxy, applied as an ensemble average to each radial bin (discussed below).

We have an effective galaxy density, $n_{\text{eff}}$, \citep{2012MNRAS.427..146H} of 12 galaxies per square arcminute. In order to minimise the contamination between the lensed galaxies and the cluster members, we only use source galaxies with a photometric redshift greater than $z_{\rm{cluster}} + 0.2$.  Our redshift cut is made so that there is negligible contamination between cluster and source galaxies.  The photo-z cut does not require a redshift dependence as the photo-z errors of the source galaxies in CFHTLenS are approximately flat close to the redshift of our clusters \citep{2012MNRAS.421.2355H}.

For each galaxy we calculate the tangential and cross shears ($\gamma_{\rm{t}}, \gamma_{\rm{x}}$) as a function of their position relative to the cluster position, via the angle $\phi$ between the cluster and galaxy from a baseline of zero declination. The tangential shear measured around each XCS determined cluster centroid was binned into 24 equal spaced logarithmic annuli out to a distance of $10\times r_{\rm{200}}$ (calculated in Section \ref{sec:Measure_X_ray}).  We then scaled the values in each of these bins in the same way that we previously scaled the X-ray profiles in Section \ref{sec:Measure_X_ray} for consistency.

Finally, in order to improve the signal to noise of the tangential profiles, the 58 individual cluster profiles were stacked.  This was achieved by summing the profiles of each cluster and calculating an average shear in each bin across all clusters (\citealt{2001astro.ph..8013M}, \citealt{2009ApJ...703.2217S}).  The error covariance matrix was then directly measured for our stacked profile.  Due to the large uncertainty in the central bin, driven by the low number density of galaxies, we exclude the central $0.1\times r_{\rm{200}}$.

We perform consistency and null tests upon the CFHTLenS shape data to ensure our recovered profiles are unbiased and not artefacts of the data.  Figure \ref{fig:null_tests}a shows the tangential signal (solid blue) and the cross shear (dashed red) around the stacked clusters.  The tangential shear signal has a detection significance of $>30\sigma$ while the cross shear signal is consistent with zero at all radii. 

Figure \ref{fig:null_tests}b shows the tangential shear (solid blue) and cross shear (dashed red) around 58 random stacked positions within the overlap of the CFHTLenS region and the XCS footprint.  The measurements in both these cases were found to be consistent with zero on all scales.  

For Figure \ref{fig:null_tests}c we show the tangential shear around the stacked clusters after we have split the source galaxies into three bins based upon their signal to noise ratio, $S/N < 20$, $20 < S/N < 40$, and $S/N > 40$, with similar redshift distributions (median redshifts of 0.85, 0.82, 0.79 respectively).  We find that the three measurements are consistent with each other as expected.

Finally Figure \ref{fig:null_tests}d shows the tangential shear around the stacked clusters with the source galaxies cut into three bins based upon their photometric redshift, $z < 0.6$, $0.6 < z < 0.8$ and $z > 0.8$.  At higher redshifts there are a smaller fraction of cluster galaxies and galaxies in front of the clusters, and the weak lensing signal grows with redshift.  We see these effects as our measured signal is strongest in the high redshift bin.  We therefore conclude we are detecting a genuine weak lensing signal.

\subsection{Binning in X-ray Temperature}
To generate tighter constraints upon the modified gravity parameters we split our data set into two separate mass bins to reduce errors caused by mixing clusters of varying sizes and masses. We find doing so improves our constraints on the modified gravity parameters compared to using a single bin.  We cut at an X-ray temperature of $T=2.5$keV, to give two bins of mass with equal errors on their stacked profiles.  We note that this temperature cut approximately cuts our sample into galaxy clusters and galaxy groups \citep{2012MNRAS.422.2213S}.  Our low temperature bin ($T<2.5$keV) has a median redshift of $z=0.32$ and is flagged with a 0 in Table \ref{table:sources}, while the other (with $T>2.5$keV) has a median redshift of $z=0.34$ and a flag of 1.  We repeated the analyses with three and four temperature bins and found no improvement in the constraints on the modified gravity parameters.  Therefore to aid with computation, we complete our analysis with the simplest two bin case.

\subsection{MCMC Analysis}
\label{sec:mcmc}

We use MCMC \citep{gilks96} to fit models to our stacked profiles.  We allow all parameters that depend upon the cluster properties to vary for each temperature bin.  This leads to a total of fourteen free parameters for the four stacked profiles (our measured weak lensing and X-ray profiles in two temperature bins) used to constrain modified gravity.  Four of these were used to model the weak lensing mass (defined in Equations \ref{eq:NFW1},\ref{eq:NFW2},\ref{eq:NFW3}).  We introduce the notation $\rm{I}, \rm{II}$ to indicate the temperature bins $T<2.5, T>2.5$ respectively so $c^{\rm{I}}$, $c^{\rm{II}}$, $M^{\rm{I}}_{\rm{200}}$ and $M^{\rm{II}}_{200}$ are the concentration and mass parameters for each temperature bin respectively.

We modelled the X-ray surface brightness, using the method prescribed in Section \ref{sec:Theory} by defining, for both temperature bins, the electron number density (itself dependent upon $n^{\rm{I}}_{\rm{0}}$, $n^{\rm{II}}_{\rm{0}}$, $b^{\rm{I}}_{\rm{1}}$, $b^{\rm{II}}_{\rm{1}}$, $r^{\rm{I}}_{\rm{1}}$ and $r^{\rm{II}}_{\rm{1}}$), and the normalisation of the gas temperature $T^{\rm{I}}_{\rm{0}}$ and $T^{\rm{II}}_{\rm{0}}$.  We reconfigure the parameters as $\beta_{\rm{2}}=\beta/(1+\beta)$ and $\phi_{\infty,\rm{2}}=1-\exp(-\phi_{\infty}/10^{-4}M_{\text{Pl}})$ to span the parameter range of $\beta$ and $\phi_{\infty}$ in the interval [0,1].  To obtain the cooling function (used in Equation \ref{eq:SBprofile}), we used the XSPEC software \citep{1996ASPC..101...17A} and utilise the APEC model \citep{2001ApJ...556L..91S} over a range of 0.5keV to 2keV, i.e. the same energy range as our observations from {\it XMM}.  This model has as inputs the gas temperature, the cluster redshift, the cluster metallicity and a normalisation, and provides the X-ray cluster flux. We adopt a metallicity $Z=0.3 Z_{\odot}$ \citep{2011PASJ...63S.991S} throughout.  Using this model we generate fluxes for a range of temperatures which are interpolated for use in our chameleon gravity model.

The chameleon parameters $\beta_{\rm{2}}$ and $\phi_{\infty,\rm{2}}$ are the same across the two bins, as the modifications to gravity should be independent of the cluster's mass.

We performed an MCMC analysis using the \textit{emcee} code \citep{2013PASP..125..306F}, which implements a Metropolis-Hastings algorithm \citep{MacKayBook}.  We minimized the goodness of fit using a $\chi^{2}$ statistic derived from joint fitting of both models (see Appendix \ref{app:goodness_of_fit}).

Our MCMC run was a parallelised implementation using 128 walkers with 10000 time steps.  We removed the first 2000 iterations as a ``burn in'' phase.

\section{Results}
\label{sec:results}

\begin{figure*}
\includegraphics[width=18cm]{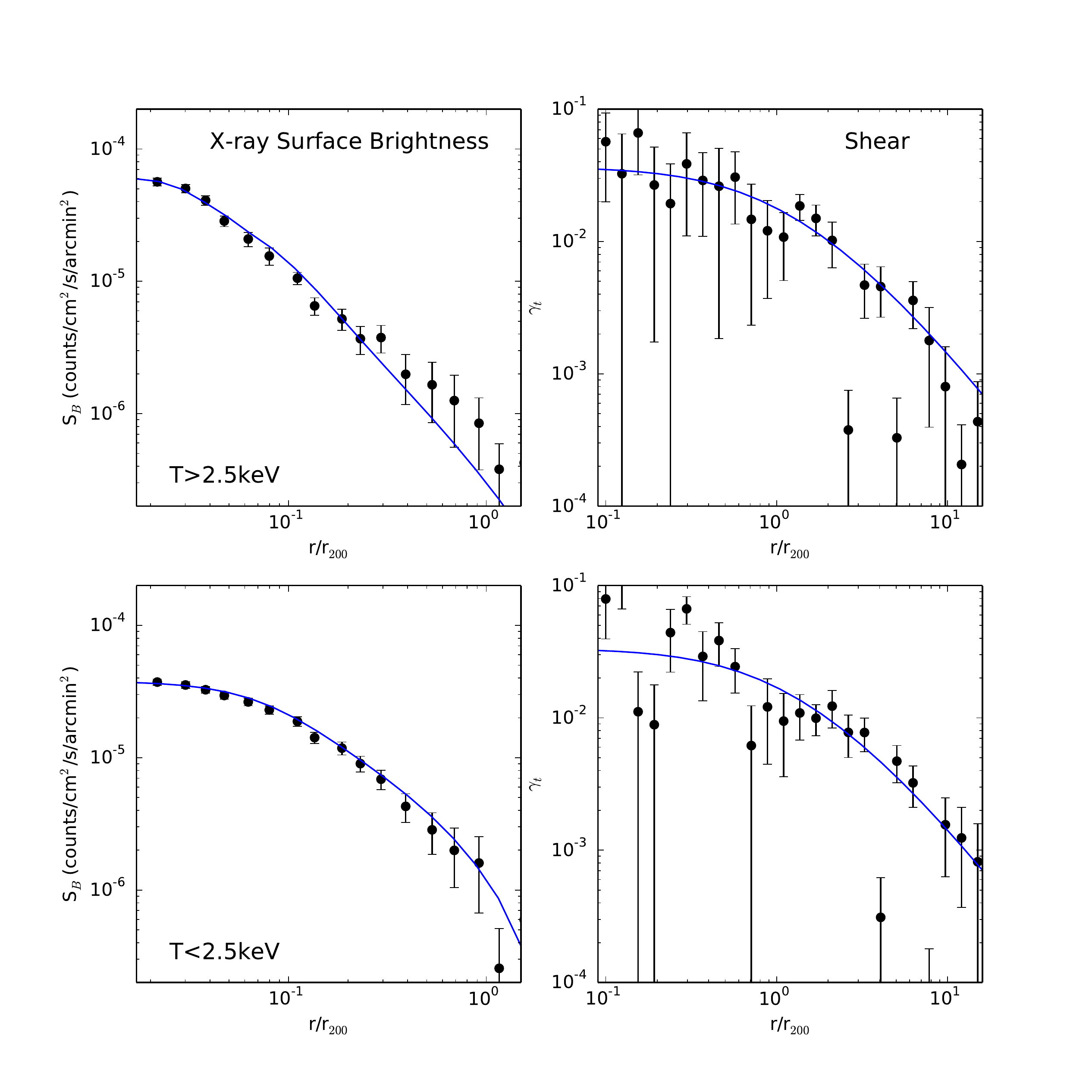}
\caption{X-ray surface brightness profiles (left) and weak lensing (right) for the two bins of X-ray temperature: $T<2.5$keV (top) and $T>2.5$keV (bottom), against radial distance normalised by $r_{200}$, the radius at which the density is two hundred times the critical density.  We choose to show the modified gravity profiles with the highest likelihood parameters, $T^{\rm{I}}_{\rm{0}}=12.6$ keV, $n^{\rm{I}}_{\rm{0}}=2.0 \times \rm{10^{-2} cm^{-3}} $, $b^{\rm{I}}_{\rm{1}}=-0.42, r^{\rm{I}}_{\rm{1}}=0.06$ Mpc, $M^{\rm{I}}_{\rm{200}} = 12.2 \times \rm{10^{14}} M_{\odot}$, $c^{\rm{I}}=3.5, T^{\rm{II}}_{\rm{0}}=7.8$ keV, $n^{\rm{II}}_{\rm{0}}=4.9 \times \rm{10^{-2} cm^{-3}} $, $b^{\rm{II}}_{\rm{1}}=-0.89, r^{\rm{II}}_{\rm{1}}=0.05$ Mpc, $M^{\rm{II}}_{\rm{200}}=13.7 \times \rm{10^{14}} M_{\odot}$, $c^{\rm{II}} = 3.8, \beta=2, \phi_{\infty}=2.1\times10^{-4}M_{\text{Pl}}$.}
\label{fig:data_fits}
\end{figure*}

\begin{figure*}
\includegraphics[width=11cm]{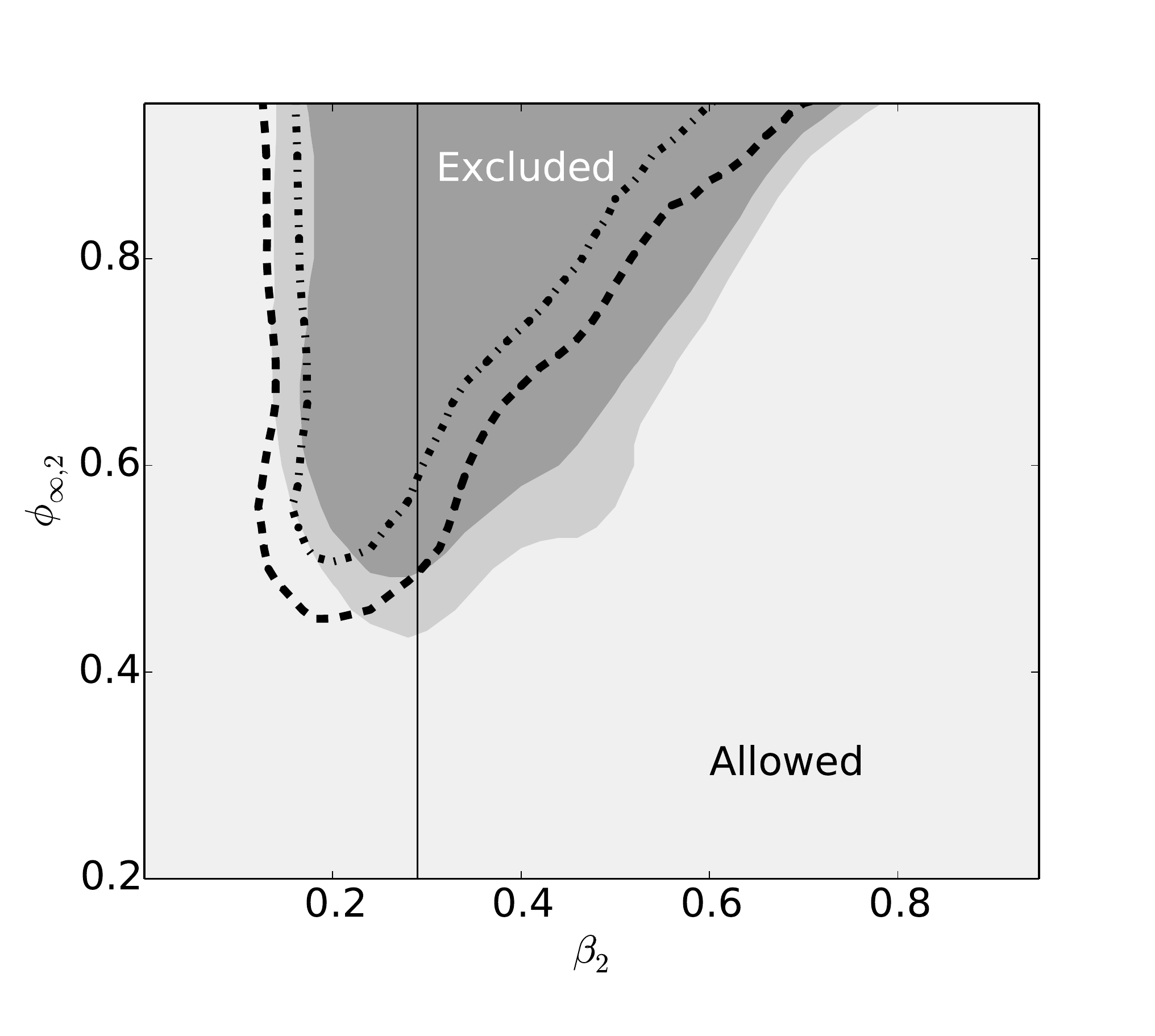}
\caption{The 95\% (light grey region) and the 99\% confidence limit (mid grey region) constraints for the chameleon model parameters renormalised between [0,1], $\beta_{\rm{2}}=\beta/(1+\beta)$ and $\phi_{\infty \rm{,2}}=1-\exp(-\phi_{\infty}/10^{-4}M_{\text{Pl}})$ obtained from the MCMC analysis of our combination of weak lensing and X-ray surface brightness for our two cluster stacks. Above the dashed (dash-dot) line is the 95\% (99\%) confidence limit excluded region from \protect\cite{2014JCAP...04..013T}.  The vertical line is at $\beta=\sqrt{1/6}$, showing our constraints for $f(R)$ gravity models.}
\label{fig:MG_contours}
\end{figure*}

In Figure \ref{fig:data_fits} we show our measured X_ray and weak lensing profiles for both X-ray temperature bins.  Our X-ray surface brightness profiles have been measured out to $1.2 \times r_{200}$ with high signal to noise.  Likewise for our two weak lensing profiles we have recovered a shear signal out to $10 \times r_{200}$ with high signal to noise.  Also shown in Figure \ref{fig:data_fits} are our best fit models for the each profile using the parameters outlined in Section \ref{sec:mcmc} and minimising $\chi^{2}$ as described in Equation \ref{eq:chi_2}. We show the 2D contours for constraints on model parameters in Figure \ref{MG_full_contours}.  

In Figure \ref{fig:MG_contours} we show the 2D constraints for $\beta_{\rm{2}}$ and $\phi_{\infty \rm{,2}}$.  To generate our constraints we have marginalised over the measured likelihoods of the nuisance parameters (those that are not $\beta_{\rm{2}}$ and $\phi_{\infty \rm{,2}}$).  We are able to do so as we are insensitive to the overall amplitude of our profiles, only the profiles shape matters for our constraints.  In Figure \ref{fig:MG_contours} we also show the dashed (dash-dot) line the 95\% (99\%) confidence limit excluded region from \cite{2014JCAP...04..013T}.  The constraints are tighter from this work on larger values of $\beta$ than in \cite{2014JCAP...04..013T}, whilst the constraints on smaller values of $\beta$ are looser.  As the profiles presented in this work extend further from the cluster than the Coma profile, we probe further outside the critical radius, $r_{\rm{c}}$ and are able to better constrain large values of $\beta$.  However, as the errors on the X-ray profiles (and the lack of available SZ data) used in this work are larger than those measured in \cite{2014JCAP...04..013T}, we are less able to differentiate a chameleon profile from a GR one at lower values of $\beta$, leading to less constraining power. 

The shape of the contours in Figure \ref{fig:MG_contours} can be understood by considering the meaning of the parameters used in defining chameleon gravity.  Recall that $\beta$ dictates the strength of the fifth force and $\phi_{\infty}$ is the effectiveness of the screening mechanism.  Therefore at low values of $\beta$, the fifth force causes a deviation to the profile which is too small to be distinguished from GR given the observational errors.  Likewise as GR gravity is recovered outside the critical radius $r_{\rm{crit}}$, this sets an upper limit on $\beta / \phi_{\infty}$.  As $\beta$ increases, a lower value for $\phi_{\infty}$ is required to keep $r_{\rm{crit}}$ within the cluster, giving rise to the triangular shape of the excluded region.

\subsection{Implications for $\lowercase{f}(R)$ Gravity}
\label{sec:f(R)}
Our constraints have implications for $f(R)$ gravity models, which contain a chameleon mechanism for which $\beta=\sqrt{1/6}$ \citep{2007JETPL..86..157S} (shown as the vertical line in Figure \ref{fig:MG_contours}).

From Figure \ref{fig:MG_contours}, we estimate an upper bound on $f(R)$ gravity of $\phi_{\infty}<5.8 \times 10^{-5} M_{\text{Pl}}$ at 95\% confidence limit, and therefore using Equation \ref{eq:f_R}, $f_{\rm{R}}(z=0.33)<4.7 \times 10^{-5}$ at 95\% confidence limit (where $z=0.33$ is our cluster samples median redshift).  The time-evolution of the background $f_{\rm{R}}(z)$ for a Hu-Sawicki follows \citep{2013MNRAS.428..743L},
\begin{equation}
f_{\rm{R}}(z) = |f_{\rm{R0}}| \frac{1}{n} [(1+3\Omega_{\Lambda})/(\Omega_{\rm{M}} (1+z)^3 + 4\Omega_{\Lambda})]^{n+1},
\end{equation}
where $n$ is a free parameter of the model.  At high redshifts, the background energy density is higher, therefore $f_{\rm{R}}(z)$ is smaller and the screening is more efficient.  So $f_{\rm{R}}(z)$ decreases by 22\% from our median redshift ($z=0.33$) to $z=0$, when $n=1$, and our constraint at $z=0$ is $|f_{\rm{R0}}| <6 \times 10^{-5}$ at 95\% confidence limit.  Considering a Hu-Sawicki model with $n=3$, our constraint becomes $|f_{\rm{R0}}| <2 \times 10^{-4}$ at 95\% confidence limit.  Our results are comparable to the results for the Coma cluster reported in \cite{2014JCAP...04..013T} of $|f_{\rm{R0}}| < 6 \times 10^{-5}$.

\section{Discussion}
\label{sec:Discussion}

In this section we discuss the influence of local overdensities upon our cluster sample.  We also question the validity of the assumptions we have made while constraining chameleon gravity, primarily the assumption that our cluster stack is in hydrostatic equilibrium.

\subsection{Influence of Cluster Environment}

\begin{figure}
\includegraphics[width=9cm]{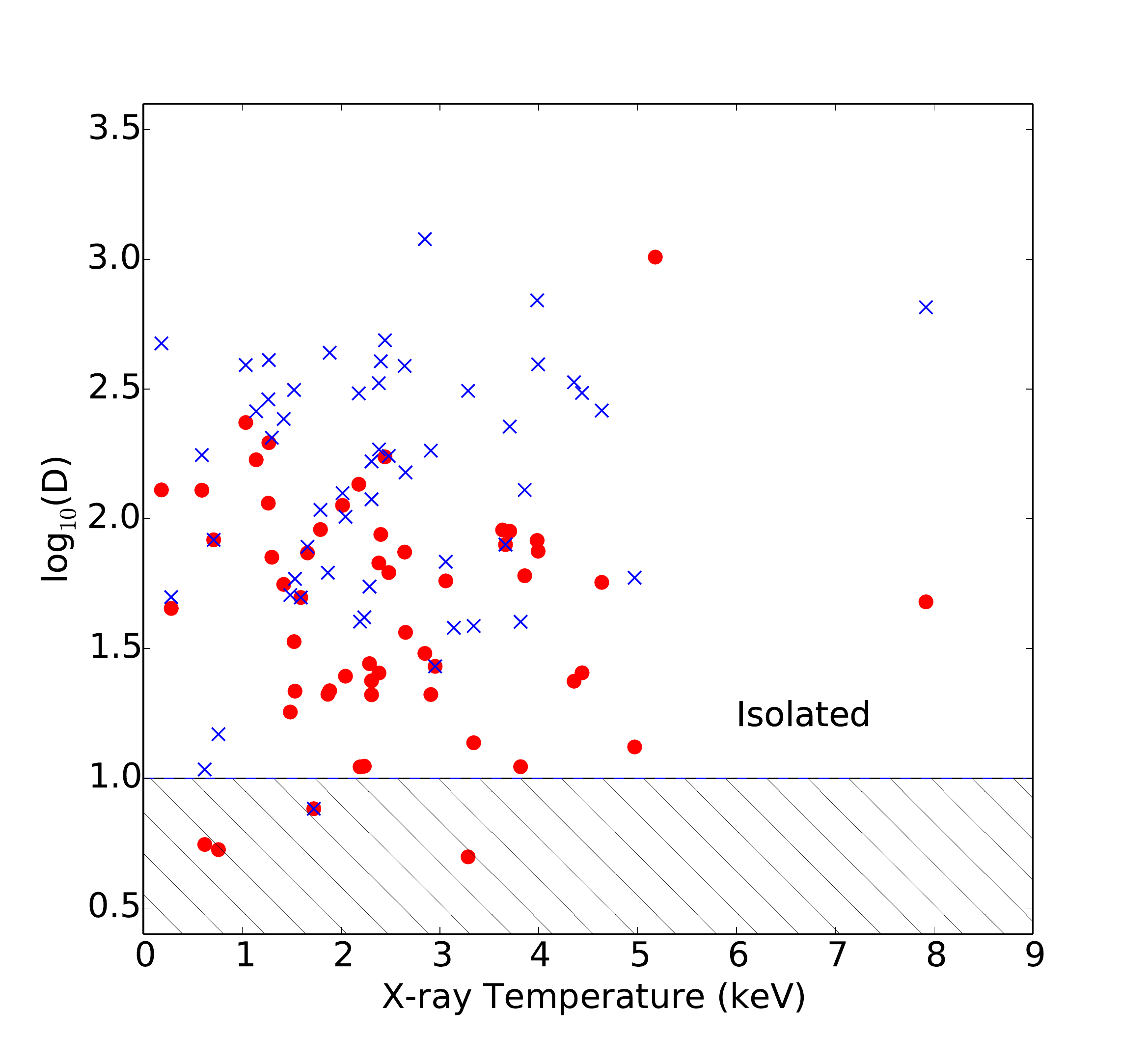}
\caption{The minimum $D$ parameter for each cluster against X-ray temperature, where $\log_{\rm{10}}D$ is a measure of the distance between a cluster and the nearest overdensity in the top 30\% (10\%) of overdensity values, shown as a red circle (blue cross).  The shaded region contains clusters with potential screening from neighbouring overdensities.  The majority of the clusters are in an isolated region.}
\label{fig:D}
\end{figure}

In addition to self screening, a cluster may be screened by nearby clusters and therefore still show no evidence of modified gravity, even in its outskirts.  To check whether this was expected for any of our clusters we estimated the $D$ parameter detailed in \cite{2011PhRvL.107g1303Z}, a parametrisation of the separation between a given cluster and the nearest larger cluster, scaled by the given cluster's $r_{\rm{200}}$.  We describe clusters with $\log_{\rm{10}}D > 1$ as ``isolated'' and clusters with $\log_{\rm{10}}D < 1$ as living in dense environments, and therefore screened.  As our X-ray clusters are an incomplete set of all clusters in our area, we looked at overdensities in the galaxy density field as a proxy for nearby clusters.  We binned the galaxies in the CFHTLenS catalogue into 3-D pixels of volume 1Mpc$^{2}$ in area, and 0.01 in redshift.  Figure \ref{fig:D} shows X-ray temperature against $\log_{\rm{10}}D$, where we have calculated $\log_{\rm{10}}D$ values between each cluster and overdensity and selected the smallest $\log_{\rm{10}}D$ as a measure of environment.  It is seen that only 7\% (2\%) of our clusters are found to be near ($\log_{\rm{10}}D<1$) the most overdense 30\% (10\%) of the 3-D pixels.  We therefore conclude that our sample appears to be largely environmentally unscreened by nearby clusters, and therefore will apply our analysis to the full cluster sample. We note that it is possible that clusters outside the edge of the CFHTLenS observations could screen at most 6\%  of our sample, which lie within $\log_{\rm{10}}D=1$ of the edge.

\subsection{Assumption of Hydrostatic Equilibrium}
\label{sec:Hydrostatic}
Even in the absence of a fifth force, the interpretation of apparent differences in cluster mass profiles derived from X-ray or Sunyaev-Zel'dovich (SZ) observations and lensing measurements is complicated by both astrophysical processes in clusters, such as gas clumping in the cluster outskirts, and systematic errors in the measurements themselves. This has led to uncertainty in mass calibration being the dominant source of error on cosmological constraints derived from SZ cluster catalogues (e.g., \citealt{2013JCAP...07..008H}; \citealt{2013ApJ...763..127R}; \citealt{2014A&A...571A..20P}). The absolute cluster mass scale is affected by uncertainty in the effects of feedback from active galactic nuclei, and non-thermal processes such as bulk motions, on the
cluster gas (e.g., \citealt{2007ApJ...668....1N}). Instrumental calibration uncertainties may also play a role (e.g., \cite{2015A&A...575A..30S}, \citealt{2015MNRAS.448..814I}). Lensing measurements, which are affected by different systematics, are being used to quantify any bias in the absolute mass scale, but at present, samples are small, and there is some disagreement (e.g., \citealt{2014MNRAS.439....2V}, \citealt{2015arXiv150201883H}).

\begin{figure*}
\includegraphics[width=18cm]{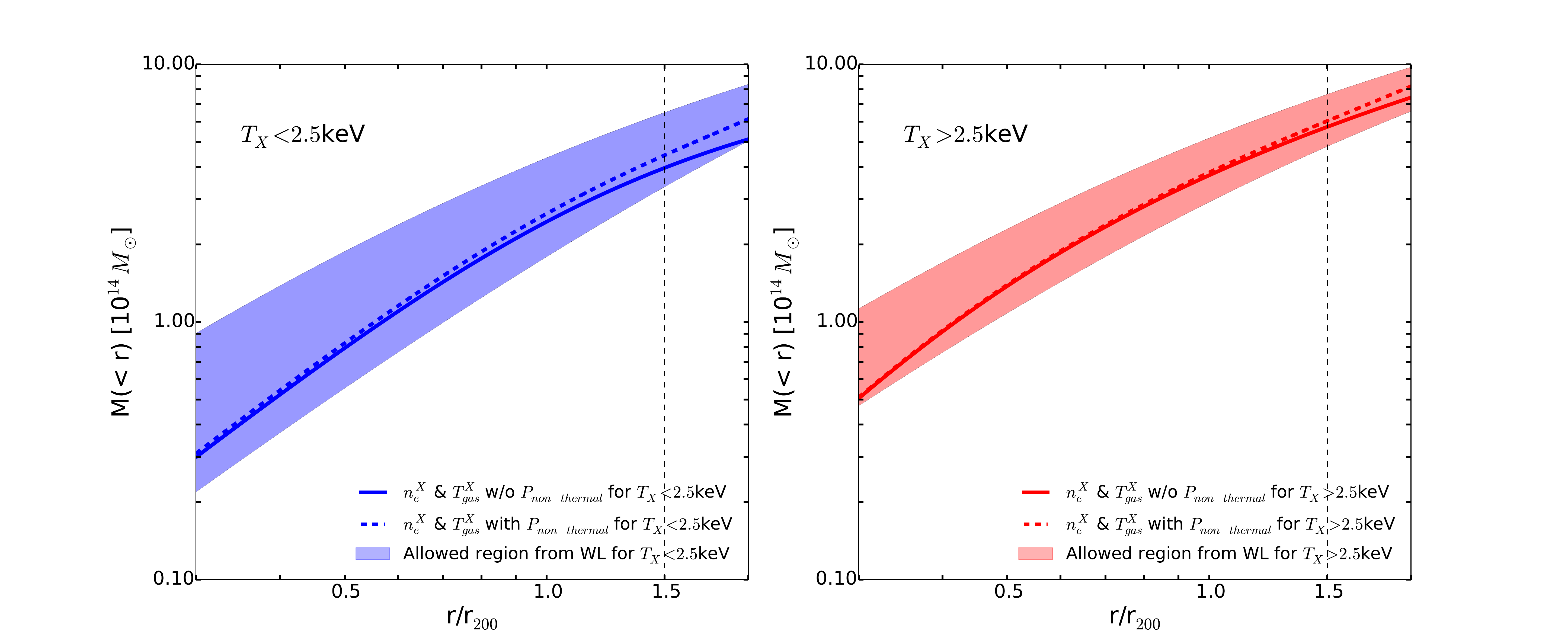}
\caption{Mass profile from the $T<2.5$keV ($T>2.5$keV) cluster bin in blue (red).  The shaded area is the one-sigma allowed region from the weak lensing measurement and the solid line is the thermal mass reconstructed from the X-rays.  The dashed line shows the thermal mass with an additional non-thermal component as discussed in the text.  The vertical line is the upper extent of our X-ray data; to its right we have extrapolated the X-ray data.}
\label{fig:hydro_mass}
\end{figure*}

In this work, we have investigated one of these issues: the impact of non-thermal pressure on our conclusions about chameleon gravity (whilst maintaining the simplifying assumptions of spherical symmetry).  We plan to investigate the other issues, using hydrodynamic simulations, in future publications.  The thermal mass of a cluster is defined by the gas pressure, density and temperature, which we infer from the X-ray surface brightness.  We follow the parametric fits described in \cite{2014JCAP...04..013T} to reconstruct the stacked cluster temperature profile and electron number densities from the profile parameters fit for by our MCMC.  We infer from X-ray observations,
\begin{equation}
M_{\rm{thermal}} = \frac{-kT_{\rm{gas}}r}{\mu m_{\rm{p}}G} \left(\frac{d \ln n_{\rm{e}}}{d \ln r} + \frac{d \ln T_{\rm{gas}}}{d \ln r} \right),
\label{eq:thermal_mass}
\end{equation}
where $k$ is the Boltzmann constant, $m_{\rm{p}}$ is the proton mass.  According to the hydrodynamical simulations in \cite{2010ApJ...725.1452S}, the non-thermal pressure can be modelled as a function of the total pressure, such that $P_{\rm{non-thermal}}(r)=g(r)P_{\rm{total}}(r)$, where
\begin{equation}
g(r)= \alpha_{\rm{nt}} (1+z)^{\beta_{\rm{nt}}} \left( \frac{\rm{r}}{r_{\rm{500}}} \right)^{n_{\rm{nt}}} \left( \frac{M_{\rm{200}}}{3 \times 10^{14} M_{\odot}} \right)^{n_{\rm{M}}},
\label{eq:g_r}
\end{equation}
with $\alpha_{\rm{nt}}$, $\beta_{\rm{nt}}$, $n_{\rm{nt}}$ and $n_{\rm{M}}$ are constants determined from 16 simulated clusters, with a mass range between $0.35-9.02 \times 10^{14} M_{\odot}$ at $z=0$ \citep{2009ApJ...705.1129L}.  We adopt their best fit values of $\beta_{\rm{nt}}, n_{\rm{nt}}, n_{\rm{M}} = 0.5,0.8,0.2$ respectively.  In order to test the robustness of our assumptions we select $\alpha=0.3$, which was the most extreme value found in the 16 clusters in their analysis.  The extra mass component which would be inferred from X-rays due to such non-thermal pressure would be
\begin{equation}
M_{\rm{non-thermal}} = \frac{-r^{2}}{G \rho_{\rm{gas}}} \frac{d}{dr} \left( \frac{g(r)}{1-g(r)} n_{\rm{gas}}kT_{\rm{gas}} \right),
\label{eq:non-thermal_mass}
\end{equation}
where $r$ is the radial distance, $g(r)$ is defined in Equation \ref{eq:g_r} and $\rho_{\rm{gas}}, n_{\rm{gas}}$ and $T_{\rm{gas}}$ are the gas density, number density and temperature respectively. 

In Figure \ref{fig:hydro_mass} we show our mass profiles for $0.3$ Mpc $< r_{\perp} < 2$ Mpc for the lensing mass and X-ray mass reconstruction, including the effects of non-thermal pressure.  The solid lines are the hydrostatic mass recovered from the X-ray measurements using Equation \ref{eq:thermal_mass}, while the dashed lines are the hydrostatic mass plus a non-thermal component from Equation \ref{eq:non-thermal_mass}. The shaded area is the 68\% confidence limit allowed region from the weak lensing measurements, fit with an NFW profile.  The vertical dotted line is the upper bound of our X-ray data; to the right of this line we have extrapolated to illustrate the possible divergence of the mass estimates with and without significant non-thermal pressure. 

At all scales in Figure \ref{fig:hydro_mass} the thermal pressure profile (solid line) is consistent with the shaded region, showing that the mass profiles estimated by the X-rays and lensing mass are consistent.  This suggests that hydrostatic equilibrium is an acceptable approximation for our stacked profiles, given the error in our lensing measurements.

We also see in Figure \ref{fig:hydro_mass} that the thermal pressure profile with a non-thermal component (dashed line) enhances the hydrodynamical mass by $20$\% ($10$\%) in the $T<2.5$keV ($T>2.5$keV) cluster bin, but is still seen to be consistent with our lensing measurements.  This shows that the non-thermal pressure expected from simulations falls within our present observed errors', if present it acts in the opposite sense to chameleon gravity, reducing the detectable signal.


With future X-ray measurements we will be able to fit out to a larger distance, allowing us to better constrain the effect of non-thermal pressure, which would be most prominent at large radii.  We also note that our weak lensing profiles have lower signal to noise than the X-ray profiles, however with future lensing surveys we will be able to more accurately constrain these profiles also allowing us to better characterise not only chameleon gravity but non-thermal pressure too.

\section{Conclusions}
\label{sec:Conclusions}
\begin{table}
\centering
\renewcommand{\arraystretch}{1.2}
\begin{tabular}{>{\arraybackslash}m{1.2in}|>{\centering\arraybackslash}m{0.4in}|>{\centering\arraybackslash}m{0.8in}}
Scale & Scale & $\log_{\rm{10}} |f_{\rm{R0}}|$ \\ \hline
Solar System \newline \citep{Hu:2007nk} & pc & $-6$ \\
Dwarf Galaxies \newline \citep{2013ApJ...779...39J} & kpc & $-6.3$ \\
Coma cluster \newline \citep{2014JCAP...04..013T} & Mpc & $-4.2$ \\
Cluster abundance \newline \citep{2014arXiv1412.0133C} & Mpc & $-4.6$ $(n=1)$ $-3.5$ $(n=3)$ \\
Cluster stack \newline (This Work) & Mpc & $-4.2$ $(n=1)$ $-3.7$ $(n=3)$ \\
CMB \newline \citep{2014PhRvD..90d3513R} & Gpc & $-3.0$
\end{tabular}
\caption{Comparison of the constraints on $\log_{\rm{10}} |f_{\rm{R0}}|$.}
\label{table:f(R)_compare}
\end{table}

We have investigated the constraining power of stacked galaxy cluster profiles for testing chameleon gravity.  We have examined 58 X-ray selected galaxy clusters, which have both good quality weak lensing data from CFHTlenS and X-ray data from XCS.  After binning our clusters by X-ray temperature, we have generated weak lensing profiles and X-ray surface brightness profiles.  Chameleon gravity predicts an additional pressure existing within clusters, which causes their gas component to become more compressed than GR gravity predicts.  We have therefore investigated this phenomena by comparing the X-ray profile with the weak lensing profile, which is unaffected by the fifth force.  Using a multi-parameter MCMC analysis we have obtained constraints on the common chameleon parameters $\beta$ and $\phi_{\infty}$, which in turn lead to constraints for $|f_{\rm{R0}}|$, a parameter charactering $f(R)$ theories. 

We find our results are competitive with other cosmological constraints on chameleon models. In particular, our constraints are an order of magnitude stronger than those from the CMB \citep{2014PhRvD..90d3513R}. They are comparable to \cite{2014arXiv1412.0133C} which provides $|f_{\rm{R0}}| < 2.6 \times 10^{-5}$ for $n=1$, compared with our measurement of $|f_{\rm{R0}}| <6 \times 10^{-5}$, and $|f_{\rm{R0}}| < 3.1 \times 10^{-4}$ for $n=3$ compared with our measurement of $|f_{\rm{R0}}| <2 \times 10^{-4}$, all at the 95\% CL.  A comparison of these constraints is shown in Table \ref{table:f(R)_compare}.

We examined the assumption of hydrostatic equilibrium by comparing the masses inferred from the X-ray observations with weak lensing and found them to be consistent. Deviations from hydrostatic equilibrium would cause a disparity between the weak lensing and X-rays with the opposite sign to that from the chameleon effect.  We modelled a non-thermal pressure X-ray component, and given current observational errors found this to be a subdominant effect on our constraints. 

As we are interested in the shape of the respective profiles, the absolute mass of the stacked cluster, measured through both weak lensing and X-rays, is a nuisance parameter which we have marginalised over. We therefore are not sensitive to the relative biases between these two techniques, such as reported in \cite{2014MNRAS.439....2V} and \cite{2015arXiv150201883H}.

For the next generation of constraints via this method, we will need detailed modified-gravity hydrodynamic-simulations.  These will allow us to check a range of assumptions used in this analysis such as hydrostaticity, non-thermal pressure, gas clumping in the cluster outskirts, spherical symmetry and the reliability of the NFW profile. 

We find our constraint on $|f_{\rm{R0}}|$ to be consistent with the literature, and competitive at these cosmic scales and redshifts.  We have therefore demonstrated that it is possible to constrain chameleon gravity using stacked galaxy clusters; with the advent of wide area lensing surveys promising a much larger area, such as the Dark Energy Survey (DES, \citealt{2005astro.ph.10346T}), the KIlo Degree Survey (KIDS, \citealt{2013ExA....35...25D}), Euclid \citep{2011arXiv1110.3193L} and the Large Synoptic Survey Telescope (LSST, \citealt{2012arXiv1211.0310L}), it will become possible to use stacks containing many more clusters to beat down systematics and obtain stronger constraints.  

\section*{Acknowledgements}
This work is based on observations obtained with MegaPrime/MegaCam, a joint project of CFHT and CEA/DAPNIA, at the Canada-France-Hawaii Telescope (CFHT) which is operated by the National Research Council (NRC) of Canada, the Institut National des Sciences de l'Univers of the Centre National de la Recherche Scientifique (CNRS) of France, and the University of Hawaii. This research used the facilities of the Canadian Astronomy Data Centre operated by the National Research Council of Canada with the support of the Canadian Space Agency. CFHTLenS data processing was made possible thanks to significant computing support from the NSERC Research Tools and Instruments grantph program.  HW acknowledges support from SEPNet, the ICG and the UK Science and Technology Facilities Council (STFC). DB, RN, AKR, KK, ARL are supported by the UK STFC grants ST/K00090/1, ST/K00090/1, STL000652/1, ST/L005573/1 and ST/L000644/1.  GBZ is supported by the 1000 Young Talents program in China, and by the Strategic Priority Research Program ``The Emergence of Cosmological Structures'' of the Chinese Academy of Sciences, Grant No. XDB09000000. JPS acknowledges support from a Hintze Fellowship.  PTPV acknowledges financial support by Funda\c{c}\~{a}o para a Ci\^{e}ncia e a Tecnologia through project UID/FIS/04434/2013.  We thank the referee for their thoughtful comments on the paper.

Numerical computations were performed on the Sciama High Performance Computing (HPC) cluster which is supported by the ICG, SEPNet and the University of Portsmouth.

\bibliographystyle{mn2e}
\bibliography{mod_grav_mnras}
\appendix
\section{Goodness of fit}
\label{app:goodness_of_fit}
To characterise the goodness of fit of our profiles we adopt the following $\chi^{2}$ statistic
\begin{align} \label{eq:chi_2}
\chi^{2}(T^{\rm{I}}_{\rm{0}}, n^{\rm{I}}_{\rm{0}}, b^{\rm{I}}_{\rm{1}}, r^{\rm{I}}_{\rm{1}}, M^{\rm{I}}_{\rm{200}}, c^{\rm{I}}, T^{\rm{II}}_{\rm{0}}, n^{\rm{II}}_{\rm{0}}, b^{\rm{II}}_{\rm{1}}, r^{\rm{II}}_{\rm{1}}, M^{\rm{II}}_{\rm{200}}, \\ \nonumber c^{\rm{II}}, \beta_{\rm{2}}, \phi_{\infty,\rm{2}}) = \chi^{\rm{I}\;\;\;2}_{\rm{WL}} + \chi^{\rm{II} \;\;\; 2}_{\rm{WL}} + \chi^{\rm{I}\;\;2}_{\rm{SB}} + \chi^{\rm{II}\;\;2}_{\rm{SB}},
\end{align}
where we adopt the notation $\rm{I}, \rm{II}$ to indicate the temperature bins $T<2.5, T>2.5$ respectively, and
\begin{equation}
\chi^{\rm{I}\;\;\;2}_{\rm{WL}}=\sum\limits_{i}\frac{(\gamma(r^{\rm{I}}_{\perp \rm{,i}})-\gamma^{\rm{obs,I}}_{\rm{i}})^{2}}{(\sigma \gamma^{\rm{obs,I}}_{\rm{i}})^{2}},
\end{equation}
\begin{equation}
\chi^{\rm{II}\;\;\;2}_{\rm{WL}}=\sum\limits_{i}\frac{(\gamma(r^{\rm{II}}_{\perp \rm{,i}})-\gamma^{\rm{obs,II}}_{\rm{i}})^{2}}{(\sigma \gamma^{\rm{obs,II}}_{\rm{i}})^{2}},
\end{equation}
\begin{equation}
\chi^{\rm{I}\;\;2}_{\rm{SB}}=\sum\limits_{i,j}(S_{\rm{B}}(r^{\rm{I}}_{\perp, \rm{i}})-S^{\rm{obs,I}}_{\rm{B,i}})C^{-1}_{i,j}(S_{\rm{B}}(r^{\rm{I}}_{\perp, \rm{j}})-S^{\rm{obs,I}}_{\rm{B,j}}),
\label{eq:SX_model}
\end{equation}
\begin{equation}
\chi^{\rm{II}\;\;2}_{\rm{SB}}=\sum\limits_{i,j}(S_{\rm{B}}(r^{\rm{II}}_{\perp, \rm{i}})-S^{\rm{obs,II}}_{\rm{B,i}})C^{-1}_{i,j}(S_{\rm{B}}(r^{\rm{II}}_{\perp, \rm{j}})-S^{\rm{obs,II}}_{\rm{B,j}}).
\end{equation}
In the weak lensing case we approximate the covariance matrix as diagonal; we find strong leading diagonals for the measured correlation matrices. For the surface brightness fits we minimise over the full covariance matrix due to the covariances that exist between bins; here $C$ is the error covariance matrix.  Then $\gamma(r_{\perp \rm{,i}})$ is the value of the lensing model at a distance $r_{\perp}$ from the clusters' centre; likewise $S_{\rm{B}}(r_{\perp, \rm{i}})$ is the value of the surface brightness model at a distance $r_{\perp}$ from the clusters centre. $\gamma^{\rm{obs}}_{\rm{i}}, S^{\rm{obs}}_{\rm{B,i}}$ are the observed shear profile and surface brightness profile respectively, while $\sigma \gamma^{\rm{obs}}_{\rm{i}}$ is the observed error on the shear profile.
\section{Source list}
\centering
\tablefirsthead{%
\hline
\multicolumn{1}{|c|}{XCS Name} &
\multicolumn{1}{c|}{z} &
\multicolumn{1}{c|}{Flag} \\
\hline}

\tablehead{%
\hline
\multicolumn{1}{|c|}{XCS Name} &
\multicolumn{1}{c|}{z} &
\multicolumn{1}{c|}{Flag} \\
\hline}
\tabletail{%
\hline}
\tablelasttail{\hline}
\bottomcaption{Sample of the extended X-ray sources in CFHTLenS footprint.  The XCS name and position are listed for all clusters.  Redshifts are provided where available.  The clusters forming the sample used throughout this work have a flag of 0 in the $T<2.5$keV bin and a flag of 1 in the $T>2.5$keV bin.  A flag of 2 denotes the source was discounted for having no measured redshift.  A flag of 3 denotes the source was discounted for having no measured X-ray temperature.  The full version of this table is provided via the online edition of the article.  An excerpt is provided above to illustrate form and content.} \label{table:sources}

\begin{supertabular}{|c|c|c|}
\hline
  XMMXCS J020045.8-064229.2 & 0.36 & 0\\
  XMMXCS J020119.0-064954.6 & 0.33 & 0\\
  XMMXCS J020232.1-073343.8 & 0.55 & 1\\
  XMMXCS J020334.3-055049.5 &  & 2\\
  XMMXCS J020359.1-055031.6 &  & 3\\
  XMMXCS J020405.2-050142.5 &  & 2\\
  XMMXCS J020428.5-070221.6 &  & 2\\
  XMMXCS J020432.7-064449.4 &  & 2\\
  XMMXCS J020514.7-045640.0 & 0.29 & 0\\
  XMMXCS J020611.4-061129.2 & 0.88 & 1\\
\hline
\end{supertabular}
\clearpage
\section{Cluster Images}
\noindent\begin{minipage}{\textwidth}
    \centering
    \captionsetup{width=17cm}
    \includegraphics[width=\hsize]{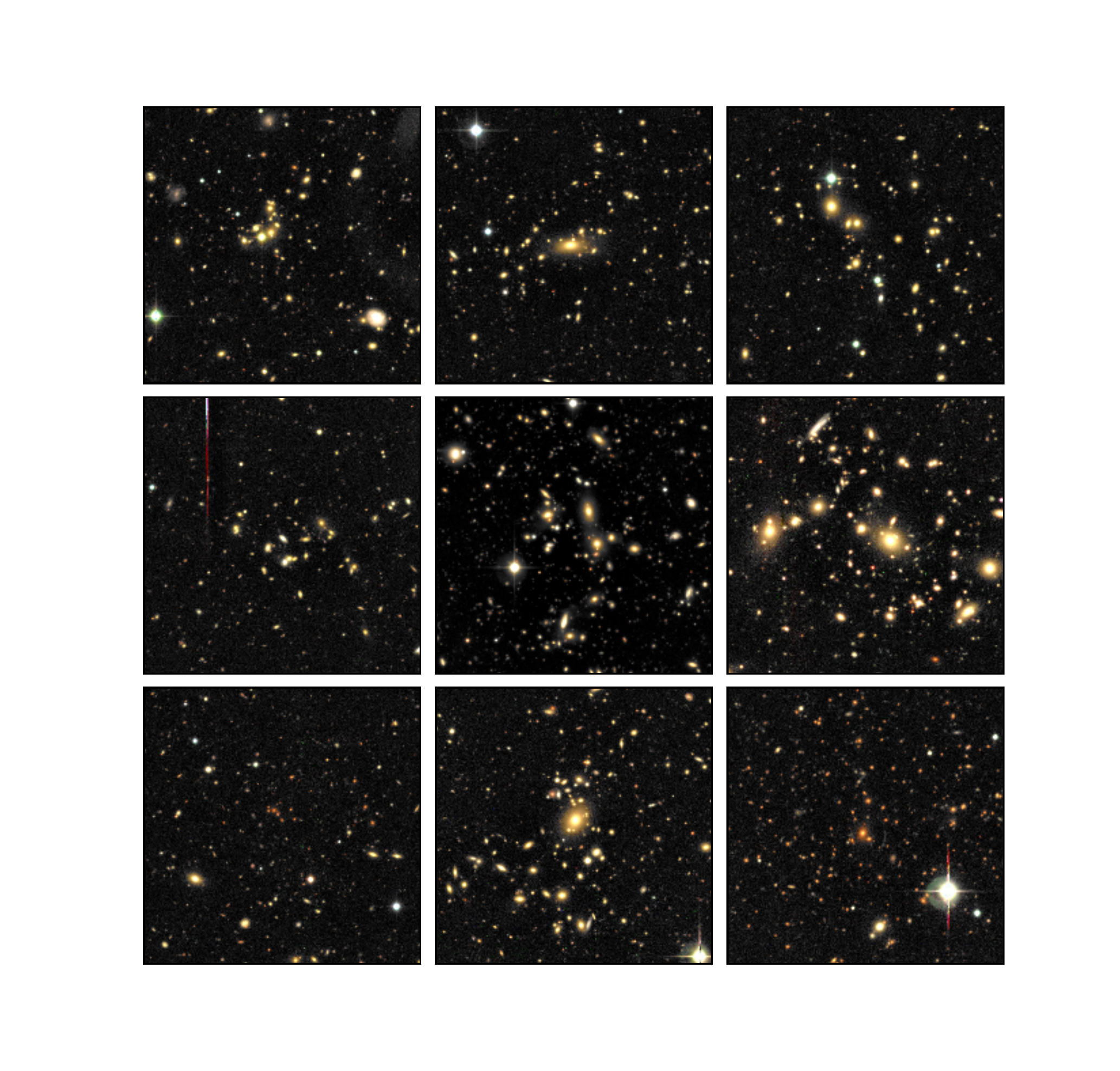}
    \label{fig:clust_pics}
    \captionof{figure}{A selection of optically confirmed clusters as imaged by CFHTLenS.  False colour composite images are $3' \times 3'$.  From left to right and top to bottom, the compilation shows the clusters: XMMXCS J020119.0-064954.6 at z=0.33; XMMXCS J021226.8-053734.6 at z=0.31; XMMXCS J021527.9-053319.2 at z=0.28; XMMXCS J021843.7-053257.7 at z=0.40; XMMXCS J022433.8-041433.7 at z=0.39; and XMMXCS J023142.2-045253.1 at z=0.21.  These clusters are included in our sample, flagged either with a 0 or 1 in Table \ref{table:sources}.  The remaining clusters in our compilation have no measured redshift or temperature and are flagged with a 2 or 3 in Table \ref{table:sources}.  Continuing onwards these clusters are: XMMXCSJ021517.1-0.60432.8; XMMXCSJ022359.2-083543.4; and XMMXCSJ141446.9+544709.1.}
\end{minipage}
\clearpage
\section{2D contours}
\noindent\begin{minipage}{\textwidth}
    \centering
    \captionsetup{width=17cm}
    \includegraphics[width=0.9\hsize]{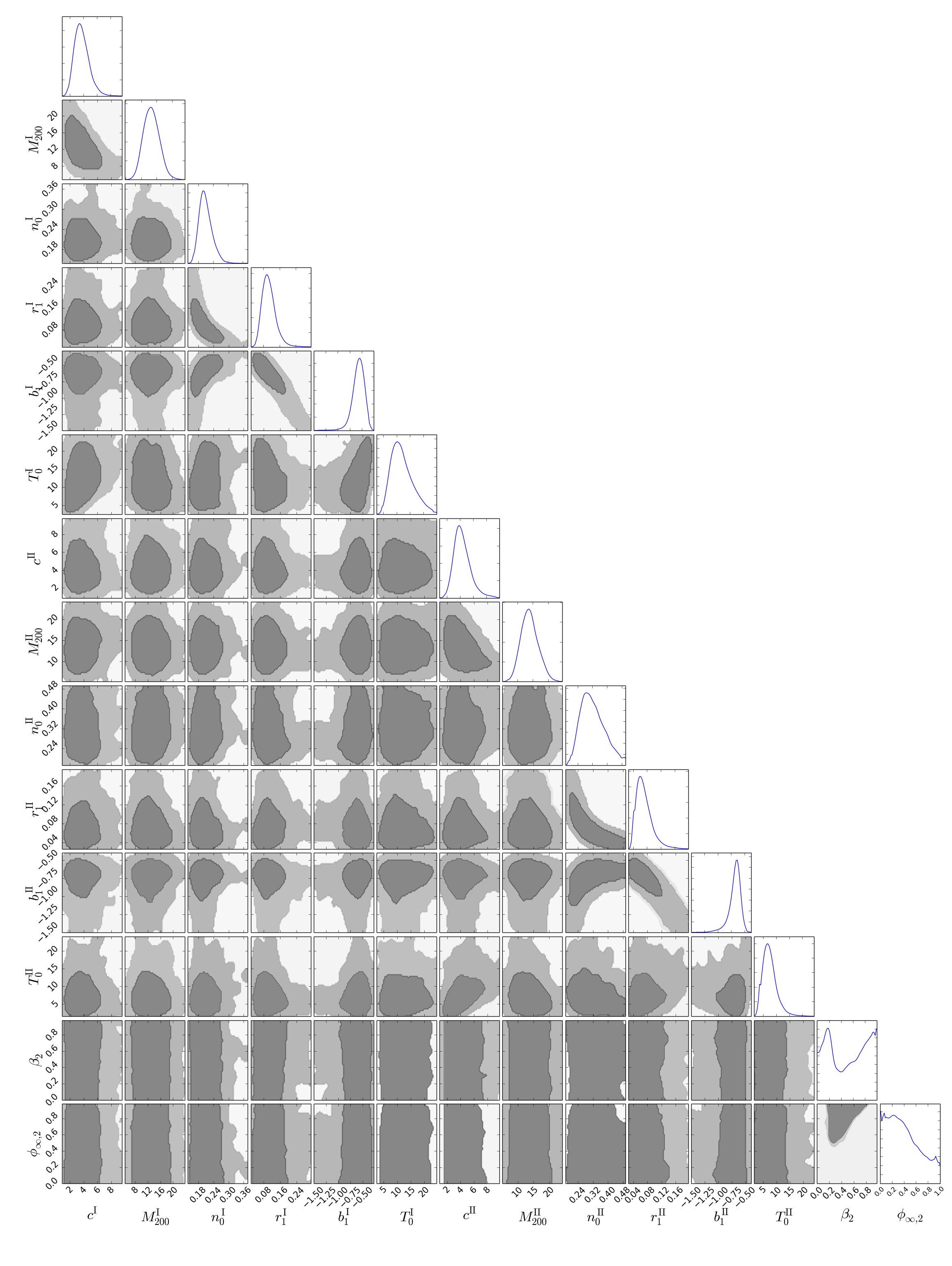}
    \label{MG_full_contours}
    \captionof{figure}{The 95\% (dark grey region) and the 99\% CL (mid grey region) 2D marginalised contours for the 14 model parameters $T^{\rm{I}}_{\rm{0}}$ [keV], $n^{\rm{I}}_{\rm{0}}$ [$\rm{10^{-2} cm^{-3}} $], $b^{\rm{I}}_{\rm{1}}, r^{\rm{I}}_{\rm{1}}$ [Mpc], $M^{\rm{I}}_{\rm{200}}$ [$\rm{10^{14}} M_{\odot}$], $c^{\rm{I}}, T^{\rm{II}}_{\rm{0}}$ [keV], $n^{\rm{II}}_{\rm{0}}$ [$\rm{10^{-2} cm^{-3}} $], $b^{\rm{II}}_{\rm{1}}, r^{\rm{II}}_{\rm{1}}$ [Mpc], $M^{\rm{II}}_{\rm{200}}$ [$\rm{10^{14}} M_{\odot}$], $c^{\rm{II}}, \beta_{\rm{2}}, \phi_{\infty \rm{,2}}$ used in our MCMC analysis.  The rightmost plots show the 1D likelihood distributions.}
\end{minipage}
\end{document}